# Light-weight sodium alanate thin films grown by reactive sputtering

M. Filippi, J.H. Rector, R. Gremaud
Department of Physics and Astronomy, Condensed Matter Physics, Vrije Universiteit,
De Boelelaan 1081,
1081 HV Amsterdam, The Netherlands

M.J van Setten
Institut für Nanotechnologie, Forschungszentrum Karlsruhe P.O. box 3640 D-76021
Karlsruhe, Germany

B. Dam
Delft University of Technology, DelftChemTech, MECS, Julianaweg 136, 2628 BL
Delft, The Netherlands

We report the preparation of sodium alanate, a promising hydrogen storage material, in a thin film form using co-sputtering in a reactive atmosphere of atomic hydrogen. We study the phase formation and distribution, and the hydrogen desorption, with a combination of optical and infrared transmission spectroscopy.

We show that the hydrogen desorption, the phase segregation and the role of the dopants in these complex metal hydrides can be monitored with optical measurements. This result shows that a thin film approach can be used for a model study of technologically relevant lightweight metal hydrides.

Complex metal hydrides are solid state hydrogen storage materials with a high capacity. Among them, sodium alanate (NaAlH$_4$) is regarded as the most promising candidate, due to its high hydrogen capacity (5.6wt %) and suitable thermodynamic stability for reversible hydrogen storage in conjunction with PEM fuel cells. [1-3]

The thermodynamic and kinetic conditions for hydrogen absorption and desorption can be improved by adding a Ti-containing dopant (metallic Ti or TiCl$_3$[4,5]). In spite of an intense research activity, the correlation between the structure and location of the Ti and its role in the catalysis (de-)hydrogenation reaction is not yet fully understood.[2,3,6] Generally, (de-)hydrogenation studies are carried out on bulk samples prepared through a complex chemical route, while the dopant is added during ball milling. Alternatively, we study the kinetics and thermodynamics of metal hydrides using thin films. Using a combinatorial technique we explore alternative metal hydride storage options.[7] This method allows a fast and efficient exploration of the thermodynamic properties and has a speed of analysis that is out of reach for bulk chemical methods. Moreover, it permits a simple and direct evaluation of phase segregation phenomena by optical spectroscopy. Finally, the dopant amount and distribution can be decided *a priori* and artificial heterostructures can be realized.

Here we report the co-sputtering of Na-Al thin films in a hydrogen reactive atmosphere. Atomic hydrogen is provided by a hydrogen atomic source which splits molecular hydrogen at a hot W filament.[8] We characterize the films with optical transmission both in the as deposited state and after high temperature desorption. We find optical signatures (in the IR and UV regions) of the formation of NaAlH$_4$, which decomposes into NaH and Al after annealing. The effect of metallic Ti doping is also explored. We observe for both the undoped and Ti-doped samples that the annealing induces a macroscopic Al segregation, which probably hinders the reverse reaction under moderate conditions. This result opens the route to the analysis of the storage properties of the sodium alanate and other light-weight complex metal hydrides, such as $LiAlH_4$, with combinatorial techniques. [3]

A schematic view of the setup used for the deposition and optical characterization is shown in figure 1. Films with a thickness of 80nm are sputtered using an RF magnetron source for Na and a DC magnetron source for Al (the sputtering ratio Na/Al is 1:1). The initial pressure in the deposition chamber is less than $10^{-10}$ mbar. The $O_2$ partial pressure, as revealed by a residual gas analyzer, is lower than $10^{-11}$ mbar with $H_2$ being the main residual gas. A $H_2$ feeding pressure of 10 mbar was used for the hydrogen atomic source,, resulting in a pressure of $1.7 \times 10^{-3}$ mbar in the deposition chamber. An additional Ar partial pressure of $1.8 \times 10^{-3}$ mbar was used for the sputtering procedure. Metallic Ti was co-sputtered at a sputtering rate Ti/Na equal to 0.02 and 0.08 respectively. After deposition, the films are transferred in UHV to the optical chamber, where the optical transmission in the range 1-6 eV is measured using an optical fiber spectrometer[8]. The Infrared transmission is measured by a Bruker Fourier Transform spectrometer. The transfer to the IR spectrometer was done in air, resulting in a partial oxidation of the film. In fact we observe a $Al_2O_3$ peak in the infrared spectrum. This oxide layer is probably protecting the film from further deterioration. We checked this hypothesis by measuring the films after different air exposure times (from 600 s up to 48 h). The XRD analysis revealed that the films are amorphous.

Figure 2 illustrates the optical transmission in the range 1-6 eV for an 80 nm Na-Al-H film grown on sapphire. The lower curve (labelled "as-deposited") represents the transmission of the as-deposited film. Such a response can arise from a mixture of metallic impurities in a material transparent in the considered range. The impurities should be dispersed to such an extent that the light (with a wavelength between 100 and 1000 nm) probes an "effective" medium i.e. the typical sizes of the inhomogeneities is much smaller than the wavelength. We modelled the spectrum using an effective medium approximation (EMA, shown in figure as a thin line), relying on dielectric functions calculated from first principles (band structure) calculations.[9] For the fit shown, we used the Bruggeman approximation considering metallic Al impurities dispersed into a matrix of NaAlH$_4$, where the effective dielectric function ($\varepsilon_{eff}$) is given by:

$$(1-f)\frac{\varepsilon_M - \varepsilon_{eff}}{\varepsilon_M + 2\varepsilon_{eff}} + f\frac{\varepsilon_{Al} - \varepsilon_{eff}}{\varepsilon_{Al} + 2\varepsilon_{eff}} = 0$$

where $\varepsilon_{Al}$ and $\varepsilon_M$ are the dielectric function of the Al impurities and of the alanate matrix respectively. The only fit parameter is the volume fraction $f$. The fit catches quantitatively the data but is not able to fully reproduce the observed features (the simulation predicts a downward curvature of the transmission between 2 eV and 4 eV). This may arise from the fact that the approximation used, is far too simple (monodispersed impurities, no percolation), and from the presence of some Na impurities as well. The obtained volume fraction for Al is 19%. Note, that a similar fit can be obtained using Na$_3$AlH$_6$ (which has an optical gap at around 6 eV[9]) or NaH (gap at 5.5 eV) as matrix materials.

Subsequently, we studied the thermal desorption using high temperature-low hydrogen pressure conditions ( a quantitative temperature programmed desorption is not available in our setup). In figure 2 we show the optical transmission of the sample annealed at 110 ºC and 1 bar of H$_2$ pressure (curve labeled as "annealed"). A large increase of the transmission is observed. Such a response can be interpreted considering a mixture of a transparent material as a matrix with metallic Al as impurity (however, in this case just a few percent). In figure 2 we also report as a thin line the Bruggeman-EMA fit obtained considering NaH as matrix material and Al as impurity. Again the Bruggeman fit (here with a Al volume fraction of 3%) fails in reproducing fully the observed wide gap which starts at 4 eV (suggesting some degree of percolation, not considered in the Bruggeman formula). Once again, just on the basis of such a fit we can not discriminate between Na$_3$AlH$_6$, NaAlH$_4$ and NaH.

To solve these identification problems, we deposited a film under the same conditions on CaF$_2$, a material with a wide window in the infrared. The infrared transmission for the as deposited film, measured *ex-situ*, is shown in figure 3. The low energy gap (where the transmission drops down to zero between 500 and 1000 cm$^{-1}$) is due to the CaF$_2$ substrate. First of all we observe no NaH related features (for NaH films the transmission drops down to zero between 1100 and 1000 cm$^{-1}$, with a first absorption band at about 1300 cm$^{-1}$). Instead, we observe absorption peaks in the region between 1200 and 1800 cm$^{-1}$ (a zoom over this region is shown as inset in figure 3). We compared our data with the IR response of bulk undoped NaAlH$_4$[10], and the resemblance (Fig. 6 of ref 10) is quite convincing. The 1620 cm$^{-1}$ peak is due to Al-H stretching, while the sharp peak at 1431 cm$^{-1}$ (not commented in ref. 10), is due to Al$_2$O$_3$ (probably formed during the exposure of the film to the air). In our case the Al-H peak appears at a lower energy if compared to ref. 10 and other works[11] (the reported frequencies range from 1650 to 1670 cm$^{-1}$). The Na$_3$AlH$_6$ phase, revealed by a peak at around 1350 cm$^{-1}$, is not observed. We also measured the IR transmission of the sample on $CaF_2$ annealed at 110 ºC and 1 bar of $H_2$ (curve labeled as "annealed"). In this case we observe the formation of NaH (as shown by the strong absorption between 1100 and 1000 $cm^{-1}$ and the shoulder at about 1300 $cm^{-1}$, highlighted in the figure). The Al-H stretching peak at 1620 $cm^{-1}$, which signals the presence of the alanate, is suppressed, while the 1431 $cm^{-1}$ peak, due to $Al_2O_3$, is still there. So, as observed in bulk samples, the high-temperature treatment decomposes the as-deposited $NaAlH_4$ into NaH (and Al).

Subsequently, we tried to reload the film at 10 bar of $H_2$ pressure. No reloading was observed, at least up the pressure used. An analysis of the optical transmission in figure 2 can help us to clarify this point. In view of the IR results, the "annealed" optical transmission spectrum (fig. 2), which was ambiguously assigned to several transparent materials, can be now considered as generated by NaH mixed with a few percent of Al (in clusters smaller than 100 nm). The higher transmission implies that the rest of the Al has *segregated into macroscopic clusters* (bigger than 1000 nm). This macroscopic segregation is probably the reason for the irreversibility of the reaction (or its shift towards much higher temperatures-pressures), as observed in the bulk. Hence, we also explored the effect of metallic Ti doping in our films. We co-sputtered Ti-Na-Al at a rate Ti/Na=0.02 (2.5% Ti) and Ti/Na=0.08 (8% Ti). The two Ti concentrations we applied are around the optimal values reported for Ti doping in bulk samples.[4,5] In figure 4 we show the optical transmission for the 2.5% Ti sample (panel a), and 8% Ti (panel b), in the as deposited (lower curve) and annealed state (110 ºC-1 bar; upper curve). The response is pretty similar to the one of the undoped sample. In the as deposited state we observe a transparent material with large metallic impurities. After annealing, we have a transparent material with small metallic impurities. In the 8% Ti film the formation of NaH after high temperature treatment is even more evident, as revealed by the sharp gap at 5.5 eV. This shows that an homogeneous metallic Ti doping (2% and 8%) is not preventing the macroscopic (1000 nm) segregation of NaH and Al in thin films and therefore is not having any beneficial effect on the reversibility of the $NaAlH_4$. In fact, in both samples we did not observe re-hydrogenation.

In summary we report the synthesis and characterization of Na-Al-H thin films, prepared via reactive sputtering. From the optical and IR transmission data we conclude that the $NaAlH_4$ phase is formed directly from the elements. A thermal treatment decomposes the alanate into NaH and Al. Both in undoped samples in samples doped with 2% or 8% metallic Ti, we observed the segregation into macroscopic Al clusters (bigger than 1000 nm) on high temperature annealing. With this simple and fast thin film procedure we are able to determine the fate of the Al and its distribution, and to explore the role of Ti dopant. The optical characterization of thin films can thus be used to explore the reaction kinetics and phase segregation phenomena in light-weight metal hydrides.

Funding by the Helmholtz initiative "FuncHy" is gratefully acknowledge. This work is financially supported by the Stichting voor Fundamenteel Onderzoek der Materie (FOM) through the Sustainable Hydrogen Programme of Advanced Chemical Technologies for Sustainability (ACTS).

**References**

[1] L. Schlapbach, A.Züttel, *Nature*, **414**, 353 (2001).

[2] F. Schüth, B. Bogdanovic, M. Felderhoff, *Chemical communications*, **20**, 2249 (2004)

[3] B. Bogdanovic, M. Felderhoff, G. Streukens, *Journal of the Serbian Chemical Society*; 74 (2), 183 (2009)

[4] B. Bogdanovic, M. Schwickardi, *Journal of Alloys and Compounds*, **253-254**, 1 (1997)

[5] C.M. Jensen, R.A. Zidan, N. Mariels, A.G. Hee, C. Hagen, *International Journal of Hydrogen Energy* **24**, 461 (1999)

[6] S. Singh, S.W.H. Eijt, J. Hout, W.A. Kockelmann, M. Wagemaker, F.M.Mulder, *Acta Materialia* **55**, 5549 (2007)

[7] R. Gremaud, C. P. Broedersz, D. M. Borsa, A. Borgschulte, H. Schreuders, J. H. Rector, B. Dam, R. Griessen, *Advanced Materials*, **19**, 2813 (2007)

[8] R.Westerwaal, M. Slaman, C. P. Broedersz, D. M. Borsa, B. Dam, R. Griessen, A. Borgschulte, W. Lohstroh, B. Kooi, G. ten Brink, K. G. Tschersich and H. P. Fleischhauer *Journal of Applied Physics* **100**, 063518 (2006)

[9] M.J. van Setten, R. Gremaud, G.A. de Wijs, G. Brocks, G. Kresse, B. Dam, R. Griessen, unpublished

[10] Mehraj-ud-din Naik, Sami-ullah Rather, Renju Zacharia, Chang Su So, Sang Woon Hwang, Ae Rhan Kim, Kee Suk Nahm. *Journal of Alloys and Compounds*, **471**, L16-L22 (2009)

[11] S. Gomes, G. Renaudin, H. Hagemann, K. Yvon, M. P. Sulic and C. M. Jensen, *Journal of Alloys and Compounds*, **390**, 305 (2005)

**FIGURE CAPTIONS**

Figure1. Schematic representation of the UHV deposition chamber and of the opical chamber with the fiber set-up.

Figure2. Optical transmission for a 80nm Na-Al-H film in the as deposited (lower curve) and annealed state (upper curve). The thin lines represent a Bruggeman-EMA fit where $NaAlH_4$ was used as a matrix material and Al (19%) as a impurity for the as deposited film . In the case of the annealed film, NaH was used as a matrix material, with 3% of Al impurity.

Figure3. Infrared transmission for a Na-Al-H/$CaF_2$ film in the as-deposited and annealed state. The inset shows a zoom over the region 1200-2000 $cm^{-1}$ for the as deposited sample, to be compared with fig. 6 of ref.10.

Figure4. Optical transmission for the 2.5% sample in the as deposited and annealed state (panel a) and for the 8% Ti (panel b)

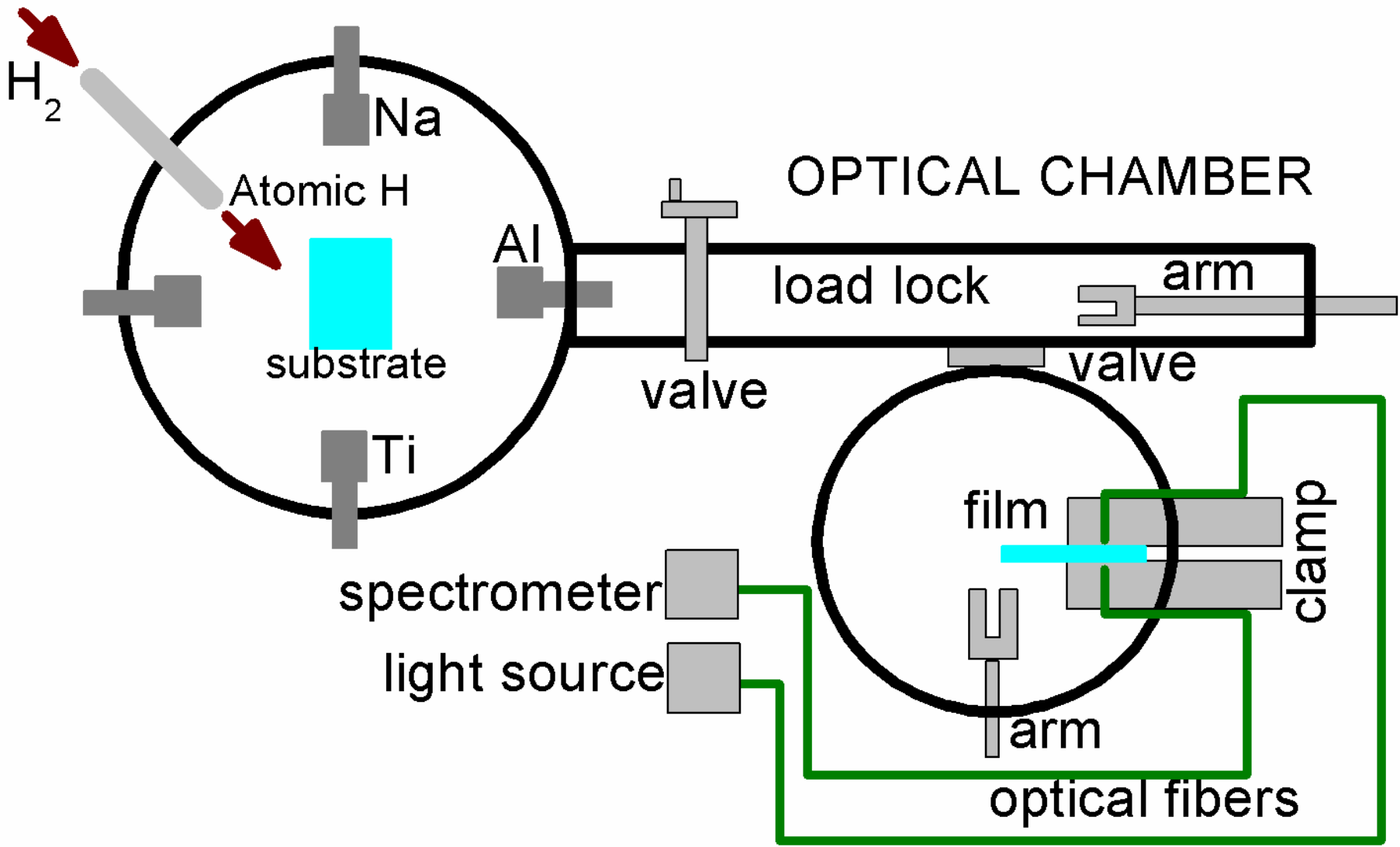

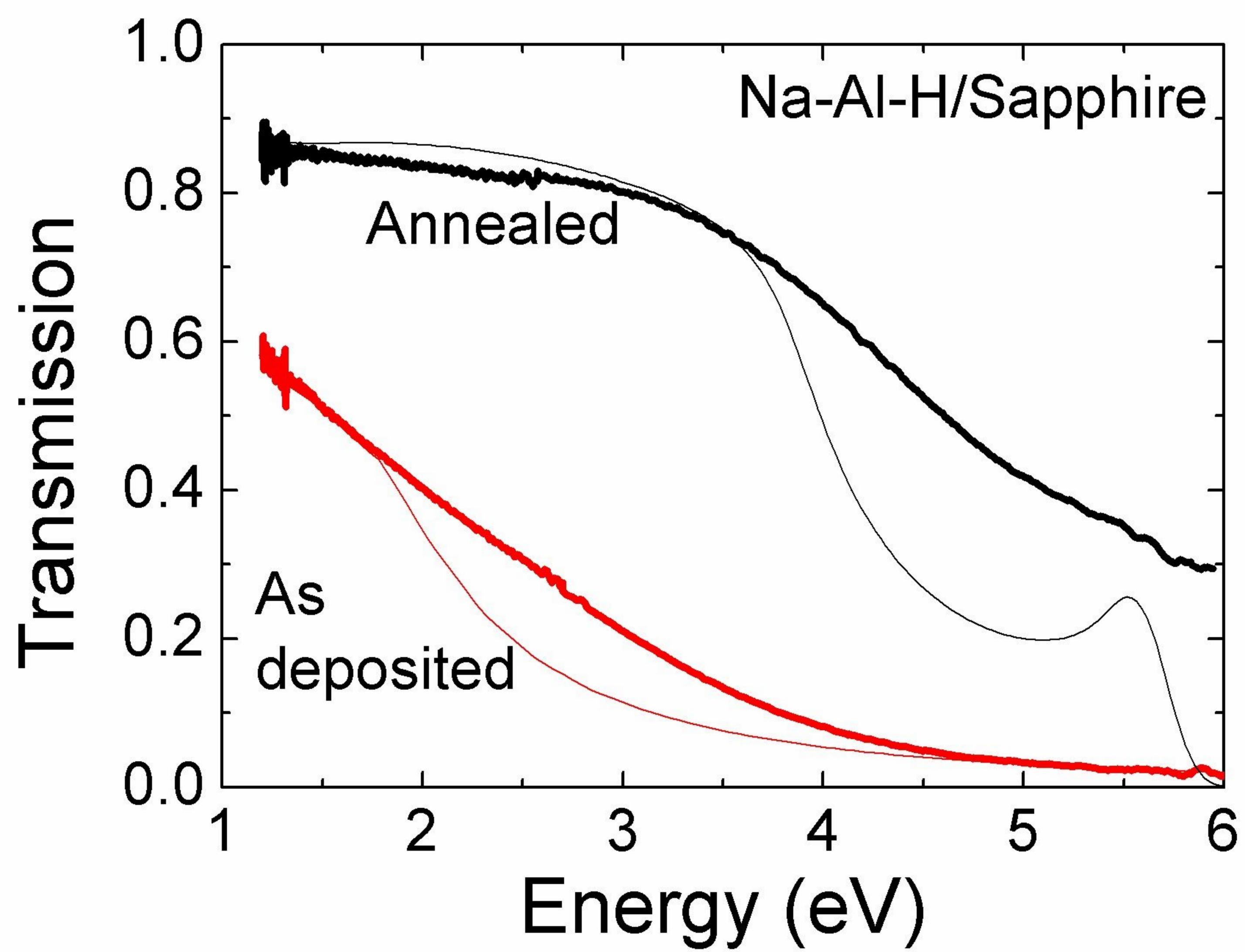

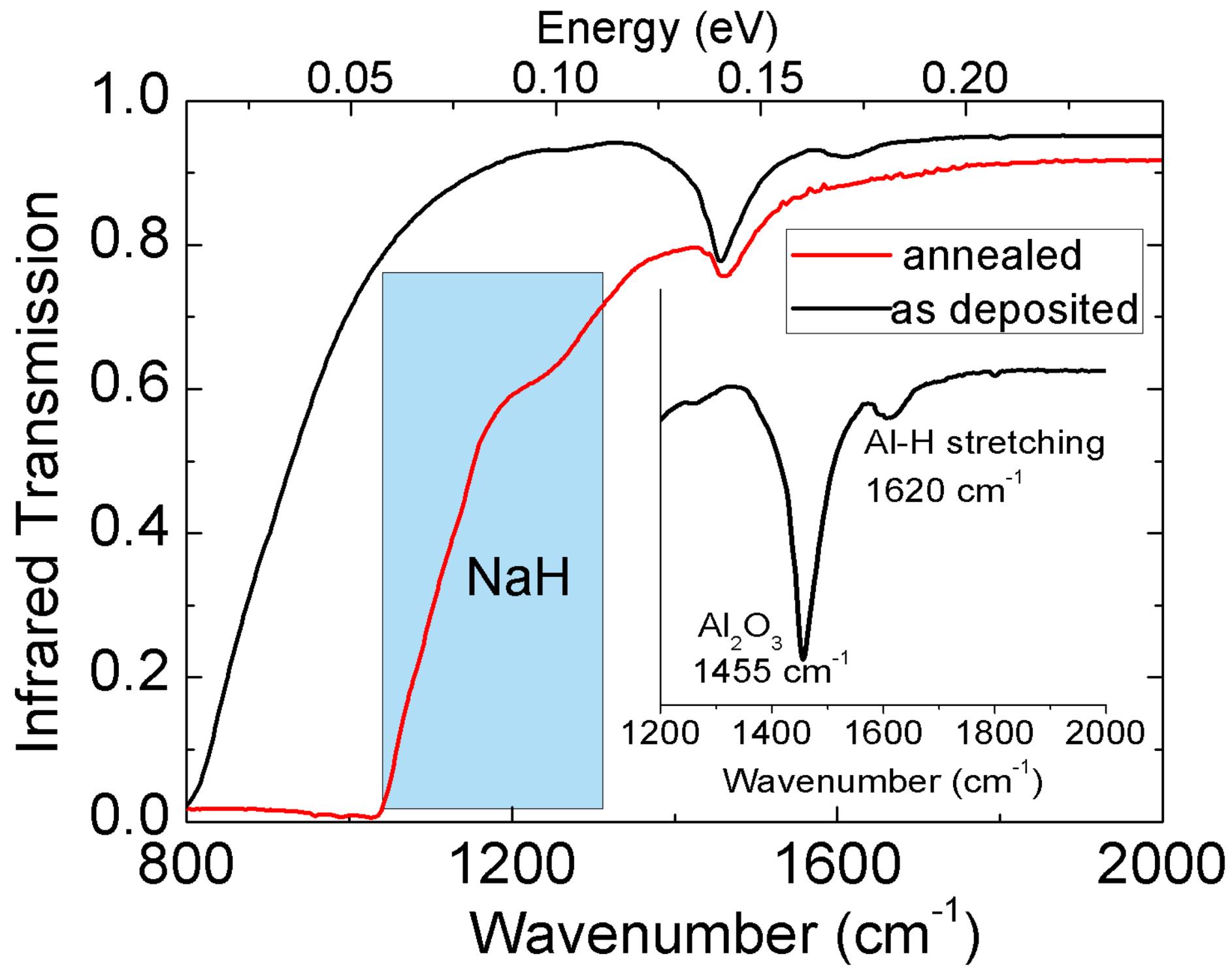

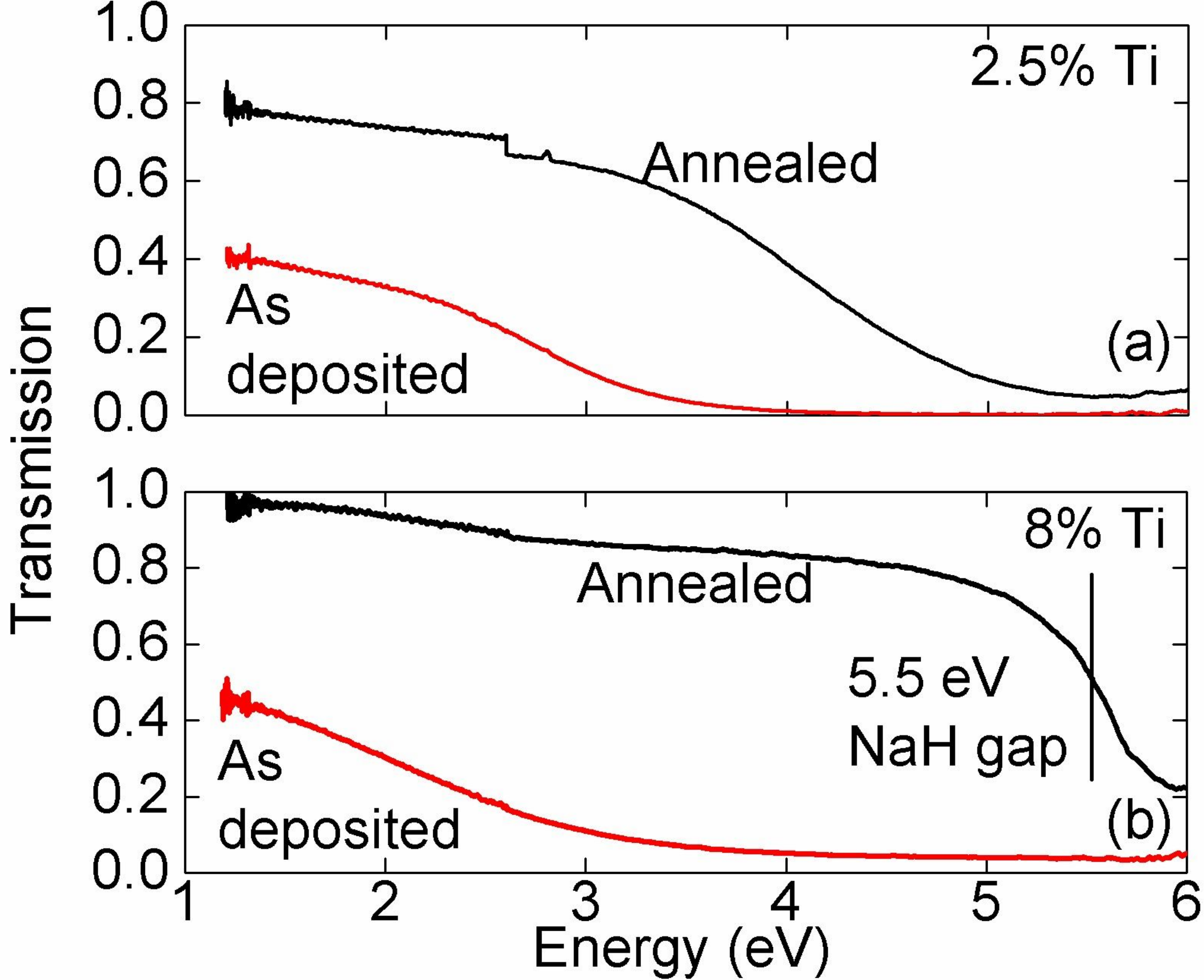